\documentclass[fleqn,10pt]{wlscirep}

\usepackage{amsmath, amssymb}
\usepackage{amsthm}
\usepackage{hyperref}
\usepackage[braket]{qcircuit}
\usepackage{caption}
\usepackage{subcaption}
\usepackage[linesnumbered,ruled,vlined]{algorithm2e}
\usepackage{tikz}
\usepackage{physics}
\usepackage[normalem]{ulem}
\usepackage{listings}
\usepackage{multirow}
\usepackage{xcolor}

\newtheorem{theorem}{Theorem}

\title{A divide-and-conquer algorithm for quantum state preparation}

\author[1,*]{Israel F. Araujo}
\author[2,**]{Daniel K. Park}
\author[3,4,+]{Francesco Petruccione}
\author[1,++]{Adenilton J. da Silva}

\affil[1]{Centro de Inform\'atica, Universidade Federal de Pernambuco, Recife, Pernambuco, Brazil}
\affil[2]{Sungkyunkwan University Advanced Institute of Nanotechnology, Suwon, Korea}
\affil[3]{Quantum Research Group, School of Chemistry and Physics, University of KwaZulu-Natal, Durban, KwaZulu-Natal, 4001, South Africa}
\affil[4]{National Institute for Theoretical Physics (NITheP), KwaZulu-Natal, 4001, South Africa}

\affil[*]{ifa@cin.ufpe.br}
\affil[**]{dkp.quantum@gmail.com}
\affil[+]{petruccione@ukzn.ac.za}
\affil[++]{ajsilva@cin.ufpe.br}

\begin{abstract}
Advantages in several fields of research and industry are expected with the rise of quantum computers. However, the computational cost to load classical data in quantum computers can impose restrictions on possible quantum speedups. Known algorithms to create arbitrary quantum states require quantum circuits with depth $O(N)$ to load an $N$-dimensional vector. Here, we show that it is possible to load an $N$-dimensional vector with exponential time advantage using a quantum circuit with polylogarithmic depth and entangled information in ancillary qubits. Results show that we can efficiently load data in quantum devices using a divide-and-conquer strategy to exchange computational time for space. We demonstrate a proof of concept on a real quantum device and present two applications for quantum machine learning. We expect that this new loading strategy allows the quantum speedup of tasks that require to load a significant volume of information to quantum devices.
\end{abstract}

\begin{document}
\flushbottom
\maketitle

\section{Introduction}

The development of quantum computers can dramatically reduce the time to solve certain computational tasks~\cite{arute2019quantum}. However, in practical applications, the cost to load the classical information in a quantum device can dominate the asymptotic computational cost of the quantum algorithm~\cite{biamonte2017quantum,Aaronson_nature2020}.
Loading information into a device is a common task in computer science applications. For instance, deep neural networks~\cite{lecun2015deep} learning algorithms run in specialized hardware~\cite{dong2019stochastic}, and the computational cost to transfer the information needs to be considered in the total computational cost as data loading can dominate the training time on large-scale systems~\cite{yang2019accelerating}. In classical devices, we can use the loaded information several times while we do not erase it. The situation is not the same in quantum devices because of the no-cloning theorem~\cite{wootters1982single}, noisy quantum operations~\cite{preskill2018quantum}, and the decoherence of quantum information~\cite{hughes}.
The no-cloning theorem shows that it is not possible to perform a copy of an arbitrary quantum state. When a quantum operation is applied, its input is transformed or is destroyed (collapsed). Even if we represent the information in a basis state that we can copy, the noisy operations and decoherence will corrupt the stored state, and it will be necessary to reload the information from the classical to the quantum device.

Loading an input vector $\vec{x}=(x_0, \cdots, x_{N-1})$ to the amplitudes of a quantum system corresponds to create the state with $\log_2(N)$ quantum bits described in Eq.~\eqref{eq:qubit}.
\begin{equation}
x_0\ket{0} + \cdots + x_{N-1}\ket{N-1}
\label{eq:qubit}
\end{equation}
Circuits to load an $N$-dimensional classical unit vector in quantum devices use $n = \log_2(N)$ qubits and have an exponential depth in relation to the number of qubits (or polynomial in the data size)~\cite{mottonen2005transformation, shende2006synthesis, iten2016quantum, park2019circuit}.

Here we propose a new format of data encoding. Namely, we load an $N$-dimensional vector in probability amplitudes of computational basis state with entangled information in ancillary qubits as
\begin{equation}
x_0\ket{0}\ket{\psi_0} + \cdots + x_{N-1}\ket{N-1}\ket{\psi_{N-1}},
\label{eq:entangledqubit}
\end{equation}
where $\ket{\psi_j}$ are unit vectors. 
We propose an algorithm to load an $N$-dimensional vector in a quantum state as shown in Eq.~\eqref{eq:entangledqubit} using a circuit with $O(\log_2^2(N))$ depth and $O(N)$ qubits. The devised method is based on quantum forking~\cite{park2019circuit,Park2019NJP} and uses a divide-and-conquer strategy~\cite{cormen2009introduction}. The circuit depth is decreased at the cost of increasing the circuit width and creating entanglement between data register qubits and an ancillary system. Thus when the data register is considered alone (i.e. by tracing out the ancilla qubits), the resulting state is mixed and not equal to the pure state shown in Eq.~(\ref{eq:qubit}). However, it is important to note that in Eq.~(\ref{eq:entangledqubit}) the classical data is still encoded as probability amplitudes of an orthonormal basis set. Useful applications can be constructed based on this, and we provide two example applications in machine learning and statistical analysis.

The divide-and-conquer paradigm is used in efficient algorithms for sorting~\cite{hoare1961algorithm}, computing the discrete Fourier transform~\cite{gentleman1966fast}, and others~\cite{cormen2009introduction}. 
The main idea is to divide a problem into subproblems of the same class and combine the solutions of the subproblems to obtain the solution of the original problem. The circuit based divide-and-conquer state preparation algorithm has computational cost $O(N)$ and the total complexity time is $O_c(N) + O_q(\log_2^2(N))$, where $O_c(N)$ is classical pre-computation time to create the quantum circuit that will load the information in the quantum device and $O_q(\log_2^2(N))$ is the depth of the quantum circuit. With the supposition that we will load the input vector $m\gg N$ times, the amortized computational time to load the real vector is $O_q(\log_2^2(N))$. The modified version of the loading problem allows an exponential advantage in the depth of the quantum circuit using $O(N)$ qubits.

The remainder of this paper is organized into 3 sections. Section~\ref{sec:tqs} reviews one of the standard methods for loading information in a quantum device using controlled rotations~\cite{mottonen2005transformation}, which we set out to modify to reduce its quantum circuit depth exponentially. Section~\ref{sec:dcmethod} shows the main result, a quantum circuit with depth $O(\log_2^2(N))$, and $O(N)$ qubits to load an $N$-dimensional vector in a quantum state with entangled information in the ancillary qubits. Section~\ref{sec:conclusion} presents the conclusion and possible future works.

\section{Transformation of quantum states}
\label{sec:tqs}

In this section, we review a strategy for loading a real vector into the amplitudes of a quantum state using a sequence of controlled one-qubit rotations~\cite{mottonen2005transformation}. Given an $N$-dimensional vector $x$, where $n= \log_2(N)$ is an integer, we can create a circuit to load this vector in a quantum computer using Algorithm~\ref{alg:ae}. The task of amplitude encoding (Algorithm~\ref{alg:ae}) has two parts: i) Function \textbf{gen\_angles} (Line~\ref{func:ga}) finds angles to perform rotations that lead $\ket{0}_n\equiv\ket{0}^{\otimes n}$ to the state in Eq.~\eqref{eq:qubit}, and ii) Function \textbf{gen\_circuit} (Line~\ref{func:gc}) generates a quantum circuit from these rotations.

\begin{algorithm}
    \SetKwInOut{Input}{input}\SetKwInOut{Output}{output}
    \SetKwFunction{genangles}{gen\_angles}
    \SetKwFunction{genanglesz}{gen\_angles\_z}
    \SetKwFunction{gencircuit}{gen\_circuit}
    \SetKwProg{Fn}{}{:}{}
    \Input{A vector $x$ with dimension $N=2^n$}
    \Output{A quantum circuit to load $x$ in the amplitudes of a quantum system}
    \BlankLine
        \Fn{\genangles{$x$}}
        { \label{func:ga} 
            \Input{A vector $x$ with dimension $N=2^n$}
            \Output{Angles to generate the amplitude encoding circuit}
            \BlankLine
            \If{$size(x) > 1$}{
                Create an auxiliary vector $new\_x$ with dimension $N/2$\\ \label{line:aux}
            	\For{$k\leftarrow 0$ \KwTo $\textnormal{length}(new\_x)-1$}
            	{
            		new\_x[k] = $\sqrt{|x[2k]|^2 + |x[2k+1]|^2}$\\
            	}
            	inner\_angles = \genangles{new\_x}\\ \label{line:recursivecall}
            	Create a vector $angles$ with dimension $N/2$\\ \label{line:angles}
            	\For{$k\leftarrow 0$ \KwTo $\textnormal{length}(new\_x)-1$}
            	{
            		\eIf{$new\_x[k] \neq 0$}
            		{
            		    \eIf{$x[2k] > 0$}
            			{
            				angles[k] = $2 \asin(\frac{x[2k+1]}{new\_x[k]})$\\
            			}
            			{
            				angles[k] = $2\pi - 2 \asin(\frac{x[2k+1]}{new\_x[k]})$\\
            			}
            		}
            		{
            			angles[k] = 0\\
            		}
            	}
            	angles = inner\_angles + angles\\ \label{line:concatenate}
            	\KwRet angles
            }
        }
        
        \BlankLine
        
        \Fn{\gencircuit{$angles$}}
        { \label{func:gc} 
            \Input{$N-1$ dimensional vector angles = \genangles{$x$}}
            \Output{Quantum circuit to load $x$ in the amplitudes of a quantum system}
            \BlankLine
            circuit = quantum circuit with $n=\log_2(N)$ qubits q[$0$], \dots, q[$n-1$] \\
            \For{$k\leftarrow 0$ \KwTo $N-2$}
            {
            	j = level(k)\\
            	index(k, j, q)\\
            	CR$_y$(angle[k], [q[0], \dots, q[j-1], q[j])\\
            	index(k, j, q)\\
            }
            \KwRet circuit
        }
        
        \BlankLine
        
        angles = \genangles{x} \\
        circuit = \gencircuit{angles} \\
        \KwRet\ circuit
        
    \caption{Amplitude encoding}
    \label{alg:ae}
\end{algorithm}

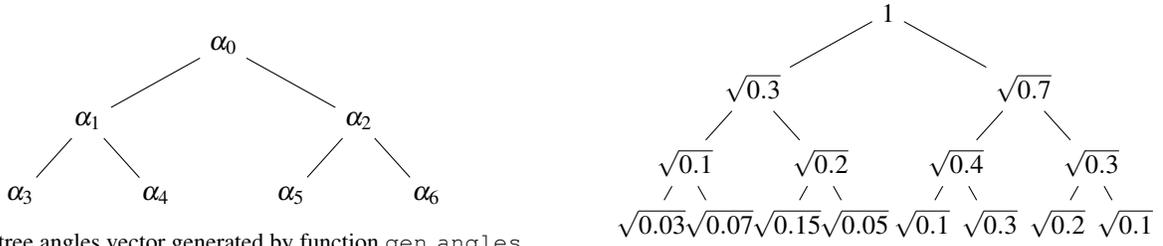
\begin{figure}[t]
\centering
\begin{subfigure}{.5\textwidth}
  \centering
\begin{tikzpicture}[level distance=1cm,
   level 1/.style={sibling distance=3.6cm, level distance=1cm},
   level 2/.style={sibling distance=1.8cm, level distance=1cm},
   level 3/.style={sibling distance=.9cm, level distance=0.8cm}]
\node {$\alpha_0$}
child{
	node{$\alpha_1$}
   child {
   node {$\alpha_3$}
	}
   child {
   node {$\alpha_4$}
   }
}
child{
node{$\alpha_2$}
   child {
   node {$\alpha_5$}
   }
child {
node {$\alpha_6$}
}
};
\end{tikzpicture}
  \caption{Binary tree angles vector generated by function \genangles with an 8-dimensional input vector.}
  \label{fig:angletree}
\end{subfigure}%
\begin{subfigure}{.5\textwidth}
  \centering
\begin{tikzpicture}[level distance=1cm,
   level 1/.style={sibling distance=3.6cm, level distance=1cm},
   level 2/.style={sibling distance=1.8cm, level distance=1cm},
   level 3/.style={sibling distance=.9cm, level distance=0.8cm}]
\node {$1$}
child{
	node{$\sqrt{0.3}$}
   child {
   node {$\sqrt{0.1}$}
   		child{node{$\sqrt{0.03}$}}
   		child{node{$\sqrt{0.07}$}}
	}
   child {
   node {$\sqrt{0.2}$}
   		child{node{$\sqrt{0.15}$}}
   		child{node{$\sqrt{0.05}$}}
   }
}
child{
node{$\sqrt{0.7}$}
   child {
   node {$\sqrt{0.4}$}
   		child{node{$\sqrt{0.1}$}}
   		child{node{$\sqrt{0.3}$}}
   }
child {
node {$\sqrt{0.3}$}
   child {
   node {$\sqrt{0.2}$}
   }
   child {
   node {$\sqrt{0.1}$}
   }
}
};
\end{tikzpicture}
  \caption{Bottom up recursive calls of the function generate angles used to compute the $\alpha$ angles of Fig.~\ref{fig:angletree}}
  \label{fig:compute_angles_example}
\end{subfigure}
\caption{Data representation of information in function generate angles.}
\label{fig:datastructuresampencoding}
\end{figure}

Function \genangles (Algorithm~\ref{alg:ae}, Line~\ref{func:ga}) divides the $2^n$-dimensional input vector into $2^{n-1}$ 2-dimensional subvectors and creates a $2^{n-1}$-dimensional vector $new\_x$ with the norms of the subvectors. While the size of $new\_x$ is greater than 1, the $new\_x$ vector is recursively passed as the input of function \genangles. This procedure is described in lines~\ref{line:aux} to \ref{line:recursivecall} of Algorithm~\ref{alg:ae}. An example of the inputs in the recursive calls with the initial input $$(\sqrt{0.03}, \sqrt{0.07}, \sqrt{0.15}, \sqrt{0.05}, \sqrt{0.1}, \sqrt{0.3}, \sqrt{0.2}, \sqrt{0.1})$$ is presented in the binary tree named state-tree in~Fig.~\ref{fig:compute_angles_example}.

After the last recursive call of the function \genangles, the algorithm starts to compute the vector angles. For each $k$ between 0 and the size of vector $new\_x$, we append an angle $\theta$ such that $\sin(\theta/2) = \frac{x[2 k+1]}{new\_x[k]}$ and $\cos(\theta/2)=\frac{x[2k]}{new\_x[k]}$ to the vector angles. Lines~\ref{line:angles} to~\ref{line:concatenate} generate the vector angles in the recursive calls. For the input in Fig.~\ref{fig:compute_angles_example} and using two decimal points the algorithm outputs angles = $(1.98, 1.91, 1.43, 1.98, 1.05, 2.09, 1.23)$. The angles vector is used as a complete binary tree named angles-tree. For instance, with $\alpha_k = angles[k]$, the angles-tree created by \genangles with an eight-dimensional input vector is described in Fig.~\ref{fig:angletree}.
Each call of \genangles will perform $\log_2(N)$ recursive calls and the cost of each call for $k = 1, \cdots,  \log_2(N)$ is $N/2^{k-1}$. The costs of the recursive calls to generate the angles-vector is $1 + 2 + 2^2 + \cdots + 2^{\log_2{N}} = O(N)$.

\begin{figure}
\center
$
\Qcircuit @C=0.22em @R=1.0em {
\lstick{\ket{0}} & \gate{R_y{(\alpha_{0})}}  & \ctrlo{1} 									 &  \ctrl{1} 							   & \ctrlo{1} 		  & \ctrlo{1} & \ctrl{1} & \ctrl{1} & \qw\\
\lstick{\ket{0}} & \qw 										& \gate{R_y(\alpha_{1})} 		 &  \gate{R_y{(\alpha_{2})}}  & \ctrlo{1} 	  & \ctrl{1} & \ctrlo{1} & \ctrl{1} & \qw\\
\lstick{\ket{0}} & \qw 										& \qw									 		 &  \qw									   & \gate{R_y(\alpha_{3})} & \gate{R_y(\alpha_{4})} & \gate{R_y(\alpha_{5})} & \gate{R_y(\alpha_{6})} & \qw\\
 }
 $
\caption{Circuit to load an 8 dimensional real vector in a quantum device.}
 \label{fig:load3qubits}
\end{figure}
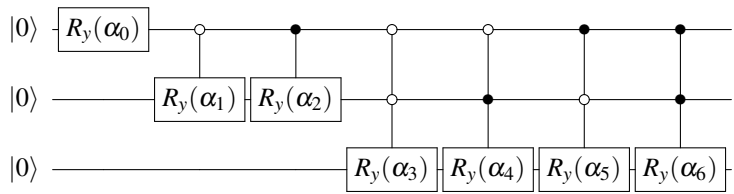

Function \gencircuit (Algorithm~\ref{alg:ae}, Line~\ref{func:gc}) receives the $N-1$ dimensional vector angles, generated by the function \genangles with input $x$, and outputs a quantum circuit to load the vector $x$ in the amplitudes of a quantum state. The state in level $j$ of the tree-state in Fig.~\ref{fig:compute_angles_example} can be constructed from the state in the level $j-1$ of the states-tree and controlled rotations from the level $j-1$ in angles-tree. The root of the angles-tree defines the first rotation and the algorithm follows a top-down approach where the rotation of angle $angle[k]$ is controlled by the qubits in range $[0, level(k))$ and is applied if the qubits $q[0], \cdots, q[level(k)-1]$ are in the state $\ket{k-(2^{level(k)}-1)}$. With $\alpha_k = angle[k]$, the circuit to load an eight-dimensional input vector is described in Fig.~\ref{fig:load3qubits}.
The computational cost to compute the angles and to generate the circuit is $O(N)$. The quantum circuit uses $O(N)$ multi-controlled gates that are applied sequentially and the circuit depth is $O(N)$. We have a $O(N)$ cost in the classical host machine and a $O(N)$ cost in the quantum device and spatial cost $O(\log_2(N))$. An amortized computational cost is $O(N)$ if we need to load the vector several times.

\section{Divide-and-conquer loading data}
\label{sec:dcmethod}
The construction of the quantum state in the previous section starts in the root of state-tree $\ket{0}_n$ and build the states in each level of the state-tree in a top-down strategy until to build the state described by the last level of the state-tree. In this Section, we propose a divide-and-conquer load strategy, and the desired quantum state is built following a bottom-up strategy. First, we divide the input into bidimensional subvectors and load qubits corresponding to the normalized bidimensional subvectors. In the next steps,  we generate the subvectors of the previous levels.

For instance, to load the state in the leafs of the state-tree in Fig.~\ref{fig:compute_angles_example}, we load four one-qubit states
$$\frac{\sqrt{0.03}}{\sqrt{0.1}}\ket{0} + \frac{\sqrt{0.07}}{\sqrt{0.1}}\ket{1}, \frac{\sqrt{0.15}}{\sqrt{0.2}}\ket{0} + \frac{\sqrt{0.05}}{\sqrt{0.2}}\ket{1}, \frac{\sqrt{0.1}}{\sqrt{0.4}}\ket{0} + \frac{\sqrt{0.3}}{\sqrt{0.4}}\ket{1} \mbox{ and } \frac{\sqrt{0.2}}{\sqrt{0.3}}\ket{0} + \frac{\sqrt{0.1}}{\sqrt{0.3}}\ket{1}$$ representing the leafs of the state-tree. To load the two two-qubit states in the previous level, the single-qubit states are weighted with the value of their fathers, obtaining the state $\ket{\psi_l}$ representing the state in the half left part of the state-tree in Eq.~\eqref{eq:leftstate} and the state $\ket{\psi_r}$ representing the state in the right part of the state-tree in Eq.~\eqref{eq:rightstate}.
\begin{equation}
\begin{split}
\ket{\psi_l}={\frac{\sqrt{0.1}}{\sqrt{0.3}}\ket{0}} \left(\frac{\sqrt{0.03}}{\sqrt{0.1}}\ket{0} + \frac{\sqrt{0.07}}{\sqrt{0.1}}\ket{1}\right) + {\frac{\sqrt{0.2}}{\sqrt{0.3}}\ket{1}} \left(\frac{\sqrt{0.15}}{\sqrt{0.2}}\ket{0} + \frac{\sqrt{0.05}}{\sqrt{0.2}}\ket{1}\right) = \\
\frac{\sqrt{0.03}}{\sqrt{0.3}}\ket{00} + \frac{\sqrt{0.07}}{\sqrt{0.3}}\ket{01} +
\frac{\sqrt{0.15}}{\sqrt{0.3}}\ket{10} + \frac{\sqrt{0.05}}{\sqrt{0.3}}\ket{11}
\end{split}
\label{eq:leftstate}
\end{equation}
\begin{equation}
\begin{split}
\ket{\psi_r}={\frac{\sqrt{0.4}}{\sqrt{0.7}}\ket{0}} \left(\frac{\sqrt{0.1}}{\sqrt{0.4}}\ket{0} + \frac{\sqrt{0.3}}{\sqrt{0.4}}\ket{1}\right) +
{\frac{\sqrt{0.3}}{\sqrt{0.7}}\ket{1}} \left(\frac{\sqrt{0.2}}{\sqrt{0.3}}\ket{0} + \frac{\sqrt{0.1}}{\sqrt{0.3}}\ket{1}\right) = \\
 \frac{\sqrt{0.1}}{\sqrt{0.7}}\ket{00} + \frac{\sqrt{0.3}}{\sqrt{0.7}}\ket{01} +
\frac{\sqrt{0.2}}{\sqrt{0.7}}\ket{10} + \frac{\sqrt{0.1}}{\sqrt{0.7}}\ket{11}
\end{split}
\label{eq:rightstate}
\end{equation}
Combining states $\ket{\psi_l}$ and $\ket{\psi_r}$ weighted with the values of the state in the previous layer generates the desired quantum state described in Eq.~\eqref{eq:finalstate}.

\begin{equation}
\begin{split}
\sqrt{0.3}\ket{\psi_l} + \sqrt{0.7}\ket{\psi_r} = & \sqrt{0.03}\ket{000} + \sqrt{0.07}\ket{001} +  \sqrt{0.15}\ket{010} + \sqrt{0.05}\ket{011} \\
& + \sqrt{0.1}\ket{100} + \sqrt{0.3}\ket{101} + \sqrt{0.2}\ket{110} +  \sqrt{0.1}\ket{111}
\end{split}
\label{eq:finalstate}
\end{equation}

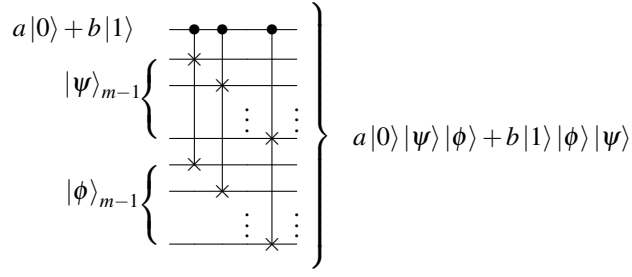
\begin{figure}[ht]
  $$
    \Qcircuit @C=0.82em @R=1.0em {
  \lstick{a\ket{0}+ b\ket{1}} &  & \ctrl{5}			& \ctrl{6} & \qw 				& \ctrl{8}    				& \qw  \\
     & & 	\qswap & \qw & \qw 				& \qw 		& \qw  \\
    & &	\qw & \qswap & \qw 				& \qw 		& \qw  \\
   & &	&  & \vdots 	&     	& \vdots  \\
   & &	\qw & \qw & \qw  	&  \qswap    	& \qw  & & \rstick{a\ket{0}\ket{\psi}\ket{\phi}+ b\ket{1}\ket{\phi}\ket{\psi}}
  {\inputgroupv{2}{5}{-0.1em}{1em}{\ket{\psi}_{m-1}}}\\
    & & \qswap	& \qw & \qw  	&  \qw   	& \qw  \\
   & &	\qw & \qswap & \qw  	&  \qw   	& \qw  \\
   & &	&  & \vdots 	&     	& \vdots  \\
   & &	\qw & \qw & \qw  	&  \qswap    	& \qw
  \inputgroupv{6}{9}{-0.1em}{1.1em}{\ket{\phi}_{m-1}}\\
  {\gategroup{1}{7}{9}{3}{1.75em}{\}}}\\
   }
   $$
   \caption{Combining states with controlled-swap operations}
   \label{fig:combining}
\end{figure}

To load the classical data using this bottom-up approach we need to combine two ($m-1$)-qubits states $\ket{\psi}, \ket{\phi}$ and one one-qubit state $a\ket{0}+b\ket{1}$ as $a\ket{0}\ket{\psi} + b\ket{1}\ket{\phi}$  with a circuit that does not depend on the input states. Using the circuit in Fig.~\ref{fig:combining} with $m-1$ controlled-swap (CSWAP) operations, we generate the desired output in the first $m$ qubits, but with unit entangled information in the $m-1$ ancillary qubits. Namely, for the example with Fig.~\ref{fig:combining}, the conventional amplitude encoding in the form of Eq.~\eqref{eq:qubit} would aim to prepare an $m$-qubit state $a|0\rangle|\psi\rangle + b|1\rangle|\phi\rangle$ while our method prepares $a|0\rangle|\psi\rangle|\phi\rangle + b|1\rangle|\phi\rangle|\psi\rangle$.

\subsection{Loading Complex data}
The divide-and-conquer strategy can be generalized to load a complex vector
$
(|x_0|e^{i\omega_0}, |x_1|e^{i\omega_1}, \dots, |x_{N-1}|e^{i\omega_{N-1}})
$
into the probability amplitudes of a quantum state as
\begin{equation}
    |x_0|e^{i\omega_0}\ket{0}\ket{\psi_0}+\dots+ |x_{N-1}|e^{i\omega_{N-1}}\ket{N-1}\ket{\psi_{N-1}}.
    \label{eq:complexsp}
\end{equation}
To explain the process, we introduce two parameters used in Ref.~\cite{mottonen_transformation_2005}
\begin{equation*}
\lambda_{j,v} = \sum_{l=0}^{2^{v-1}-1} (\omega_{(2j-1)2^{v-1}+l} - \omega_{(2j-2)2^{v-1}+l})/2^{v-1}\text{ and }
\beta_{j,v} = \sqrt{\sum_{l=0}^{2^{v-1}-1} |x_{(2j-1)2^{v-1}+l}|^2}/ \sqrt{\sum_{l=0}^{2^{v}-1}|x_{(j-1)2^v+l}|^2},
\end{equation*}
where $v=1,2,\dots,n$ is the level of the tree in reverse order (i.e. $1$ for the leaf nodes and $n$ for the root node) and $j=1,2,\dots,2^{n-v}$ is the qubit index in the layer $v$. Next, one needs $N/2$ one-qubit states corresponding to the leaf nodes of the state-tree (see Fig. \ref{fig:compute_angles_example} for example) to be prepared as
\begin{equation} \label{eq:leaf_nodes}
    \ket{\psi_{j,1}}=e^{-i\frac{\lambda_{j,1}}{2}}\sqrt{1-|\beta_{j,1}|^2}\ket{0} + e^{i\frac{\lambda_{j,1}}{2}}\beta_{j,1}\ket{1}.
\end{equation}

To load the states in the previous levels (represented by $v$ on the expression below), the states of the current level ($v-1$, since $v$ is in reverse order) are weighted with the values of their parents, obtaining the state
\begin{equation}
\label{eq:combine_children}
    \ket{\psi_{j,v}}=e^{-i\frac{\lambda_{j,v}}{2}}\sqrt{1-|\beta_{j,v}|^2}\ket{0}\ket{\psi_{2j-1,v-1}} + e^{i\frac{\lambda_{j,v}}{2}}\beta_{j,v}\ket{1}\ket{\psi_{2j,v-1}}.
\end{equation}
After recursively updating the state $\ket{\psi_{j,v}}$ for $v=2, \dots, n$ and $j=1,2,\dots.2^{n-v}$, the desired quantum state is generated as
\begin{equation}
    \ket{\psi_{1,n}}=|x_0|e^{i\omega_0}\ket{0}+|x_1|e^{i\omega_1}\ket{1}+\dots+ |x_{N-1}|e^{i\omega_{N-1}}\ket{N-1}.
\end{equation}

Combining two states at children nodes in the state-tree as shown in Eq.~(\ref{eq:combine_children}) is done with controlled-swap operations as explained in the previous section, and we will need $N$ qubits with entangled auxiliary qubits to generate the state in Eq.~\eqref{eq:complexsp}. Thus the only modification in the quantum circuit is the introduction of the $R_z(\lambda_{j,v})$ rotations to set the phases, following the $R_y$ rotations. The pseudocode for generating the angles for the $R_z$ rotations is given in Algorithm~\ref{alg:gen_angles_z}.
\begin{algorithm}
    \SetKwInOut{Input}{input}\SetKwInOut{Output}{output}
    \SetKwFunction{genanglesz}{gen\_angles\_z}
    \SetKwProg{Fn}{}{:}{}

        \Fn{\genanglesz{$x$}}
        { \label{func:ga2} 
            \Input{A vector $x$ of dimension $N=2^n$ containing the phases of the input vector.}
            \Output{Phases to generate the amplitude encoding circuit}
            \BlankLine
            \If{$size(x) > 1$}{
                Create an auxiliary vector $new\_x$ with dimension $N/2$\\ \label{line:aux2}
            	\For{$k\leftarrow 0$ \KwTo $\textnormal{length}(new\_x)-1$}
            	{
            		new\_x[k] = $(x[2k]+x[2k+1])/2$\\
            	}
            	inner\_angles\_z = \genanglesz{new\_x}\\ \label{line:recursivecall2}
            	Create a vector $angles\_z$ with dimension $N/2$\\ \label{line:angles2}
            	\For{$k\leftarrow 0$ \KwTo $\textnormal{length}(new\_x)-1$}
            	{
    				angles\_z[k] = $x[2k+1]-x[2k]$\\
            	}
            	angles\_z = inner\_angles\_z + angles\_z\\ \label{line:concatenate2}
            	\KwRet angles\_z
            }
        }
        
    \caption{Generate angles for $R_z$ rotations}
    \label{alg:gen_angles_z}
\end{algorithm}

\begin{figure}[ht]
\centering
\includegraphics[scale=0.83]{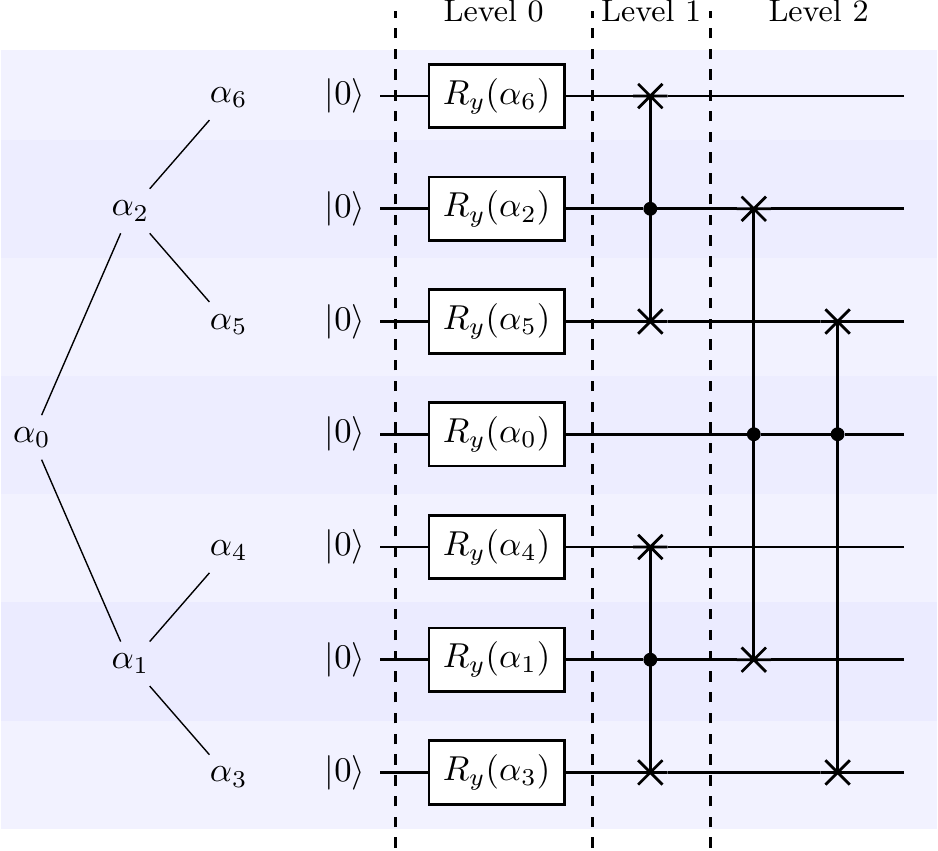}
\caption{Rotated angle-tree and a circuit generated by the divide-and-conquer strategy described in Algorithm~\ref{alg:dcload}. The quantum bit $q[k]$ in the circuit is aligned with the angle $\alpha[k]$ in the angle-tree, this organization allows to draw the quantum gates in each layer in parallel. In this example, the desired state is stored in qubits $q[0]$, $q[1]$ and $q[3]$ to generate the quantum state with entangles ancilla as in Eq.~\eqref{eq:entangledqubit}. }
\label{fig:dccircuit}
\end{figure}

Algorithm~\ref{alg:dcload} presents the complete pseudocode for the divide-and-conquer state preparation algorithm. The for loop in line~\ref{line:alg2_for} initializes the qubit $q[k]$ with the value $R_y(\alpha_k)$.  After this step, the qubits with index $k > \lfloor(N-1)/2\rfloor$  (in the leaf of the angle tree) are normalized versions of the states in the leafs of the state-tree. The next subroutine with $R_z$ rotations (Line \ref{line:rotrz_begin} to Line \ref{line:rotrz_end}) is used to encode phase information. Line \ref{line:actual} calculates the index of the first angle that has a right children in the angle-tree data structure. The while loop starting at line \ref{line:alg2_while} combines the states generated in the subtree rooted in the angle $\alpha_{actual}$. To combine the states, we first apply a cswap(q[actual], q[left\_child], q[right\_child]), and then we update the values of left and right child with the value of their left child and apply another cswap(q[actual], q[left\_child], q[right\_child]) while the left\_child and right\_child have valid values. With the input described by the angle-tree in Fig.~\ref{fig:angletree}, Algorithm~\ref{alg:dcload} generates the circuit described in Fig.~\ref{fig:dccircuit}.

\begin{algorithm}
\SetKwInOut{Input}{input}\SetKwInOut{Output}{output}
\Input{$N-1$ dimensional vector angle = \genangles{$abs(x)$}}
\Input{$N-1$ dimensional vector angle\_z = \genanglesz{$phase(x)$}}
\Output{Quantum circuit to load $x$ in the amplitudes of a quantum system entangled with ancillary qubits}
\BlankLine
circuit = quantum circuit with $N-1$ qubits q[$0$], \dots, q[$N-2$] \\
\label{line:rotry} 
\For{$k\leftarrow 0$ \KwTo $N-2$\label{line:alg2_for}}
{
	R$_y$(angle[k], q[k])\\
}
\label{line:rotrz} 

\For{$k\leftarrow 0$ \KwTo $N-2$\label{line:rotrz_begin}}
{
    R$_z$(angle\_z[k], q[k])\\
}\label{line:rotrz_end} 

actual = parent($N-2$)\label{line:actual}\\

\While{$actual \ge 0$\label{line:alg2_while}}
{
	left\_index = left(actual) \\
	right\_index = right(actual) \\
	\While{$right\_index < N-1$}
	{
		cswap(q[actual], q[left\_index], q[right\_index]) \\
		left\_index = left(left\_index) \\
	    right\_index = left(right\_index) \\
	}
	actual = actual - 1
}
\caption{Divide-and-conquer load circuit}
\label{alg:dcload}
\end{algorithm}

The process to load each state in the same layer of the state tree can be performed in parallel, because the control swap gates use different qubits. The controls are qubits in one layer of the angle-tree and targets are qubits in their subtrees.
Layer with height $k$ contributes to the depth of the circuit with the tree height minus height of the layer. The circuit will have a depth of $O(1 + 2 + \cdots + \log_2(N)-1)$ with an overall depth in order $O(\log_2^2(N))$. This result is stated in Theorem 1.

\begin{theorem}  
Algorithm~\ref{alg:dcload} generates a quantum circuit with depth $O(\log_2^2(N))$.
\end{theorem}

\subsection{Orthonormal ancillary}
\label{sec:orthonormalancillary}

The ancillary states $\ket{\psi_0}, \dots, \ket{\psi_{N-1}}$ in  Eq.~\eqref{eq:entangledqubit} are not necessarily orthogonal to each other, but we can modify the divide-and-conquer state preparation adding label qubits to ensure orthonormality of the ancillary states with the addition of label quantum register with $\log_2(N)$ qubits. The label register is prepared in $|0\rangle^{\otimes\log_2(N)}$, and $\log_2(N)$ controlled-NOT (CNOT) gates are applied to the label qubits, each controlled by a data qubit. With this modification, the final state becomes
\begin{equation}
x_0\ket{0}\ket{\psi_0}\ket{0} + \cdots + x_{N-1}\ket{N-1}\ket{\psi_{N-1}}\ket{N-1} = \sum_{k=0}^{N-1} x_k\ket{k}|\tilde{\psi}_k\rangle,
\label{eq:orthonormal}
\end{equation}
where $\lbrace |\tilde\psi_k\rangle\rbrace_{k=0}^{N-1} = \lbrace\ket{\psi_k}\ket{k}\rbrace_{k=0}^{N-1}$ is a set of orthonormal states.

\subsection{Experiments}

To evaluate the proposed method we perform two sets of experiments. 
In the first set of experiments, we use a quantum computing simulator and a NISQ computer to show as a proof of concept that the proposed method can be applied in near future. 
In the second set of experiments, we compare the depth of the circuits generated by the proposed method and other state preparation algorithms~\cite{mottonen2005transformation, shende2006synthesis} with a random input. 

\subsubsection{Proof of concept with a NISQ device}

In this experiment we load a four-dimensional data into a two qubit state $\ket{\psi} = \sqrt{0.6}\ket{00} + \sqrt{0.2}\ket{01} + \sqrt{0.1}\ket{10} +\sqrt{0.1}\ket{11}$ in a NISQ device as a proof of concept. For this experimental validation, we chose dimension of data to be small to be compatible with currently available quantum devices, although the time advantage of the proposed method will manifest when a large number of qubits are required for loading high-dimensional data.
We use qubits 1, 2 and 3 of the ibmq\_rome device. The CNOT error rates were 8.832e-3 (qubits 1 and 2) and 8.911e-3 (qubits 2 and 3). The single-qubit error was in the order of 1e-4. 

\begin{figure}[t]
\centering
    \begin{subfigure}[b]{.48\textwidth}
        \centering
        \includegraphics[width=0.95\textwidth]{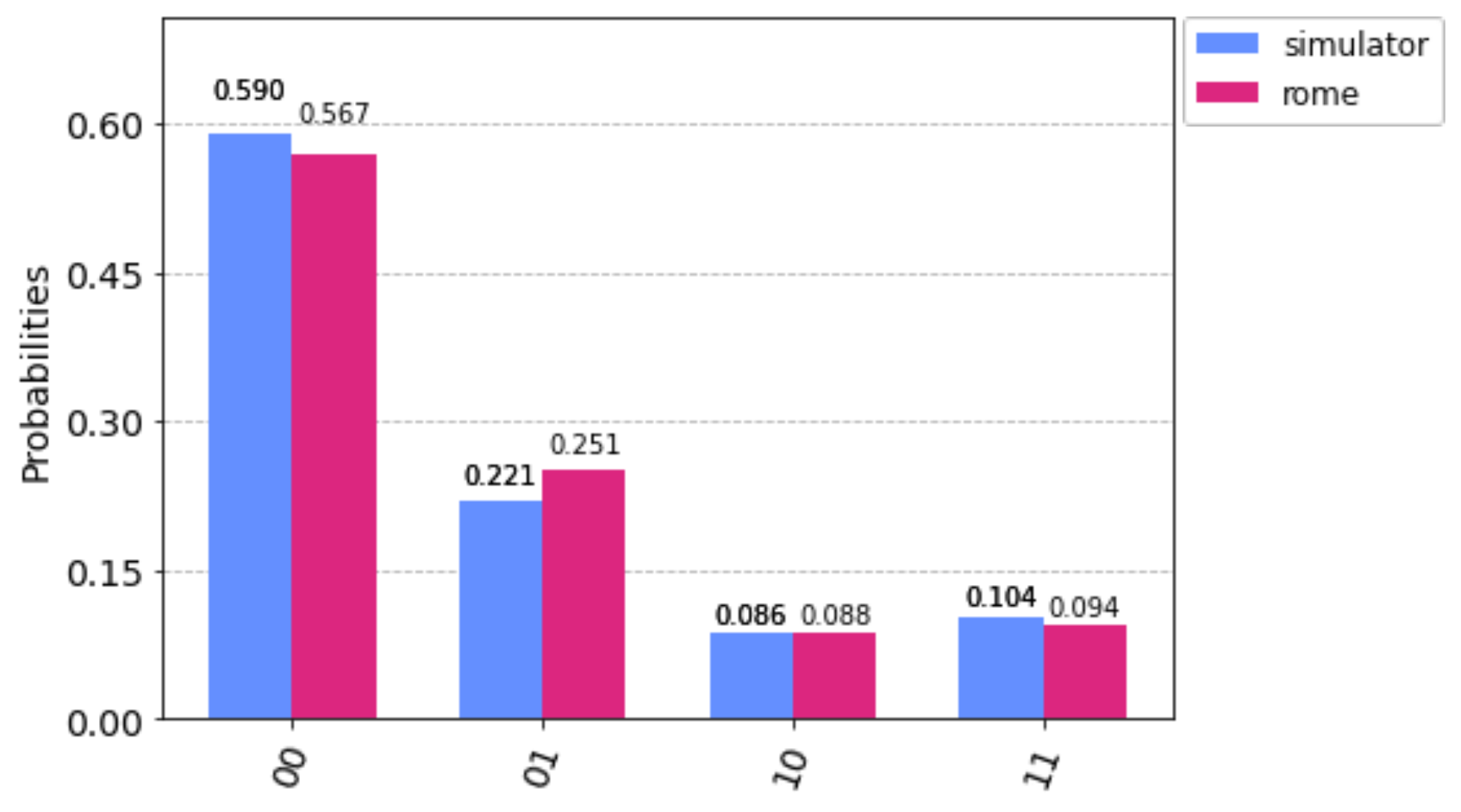}
        \caption{Output of the circuit in Fig.~\ref{fig:circ_exp} with 1024 runs \\ in a simulator (blue) and ibmq\_rome (red).}
        \label{fignisq}
    \end{subfigure}
    \begin{subfigure}[b]{.48\textwidth}
        $$
    	\Qcircuit @C=1em @R=.7em {
    	    \lstick{\ket{0}} & \gate{R_y(0.927)}   & \qw            & \ctrl{1}  & \qw       & \meter & \cw & c[1] \\
    	    \lstick{\ket{0}} & \gate{R_y(\pi/3)}   & \ctrl{1}       & \targ     & \meter    & \cw  & \cw & c[0]       \\
    	    \lstick{\ket{0}} & \gate{R_y(\pi/2)}   & \targ          & \ctrl{-1} & \qw          &        
    	}
    	$$
    	\caption{Circuit to prepare the quantum state $\ket{\psi} = \sqrt{0.6}\ket{00} + \sqrt{0.2}\ket{01} + \sqrt{0.1}\ket{10} +\sqrt{0.1}\ket{11}$ with an ancilla qubit before compilation. The measurement outcomes are stored in classical registers denoted by $c[0]$ and $c[1]$.}
        \label{fig:circ_exp}
    \end{subfigure}
    \caption{Proof of concept experiment with  a IBM quantum device (ibmq\_rome) on the cloud platform.}
\end{figure}

Figure~\ref{fignisq} presents the output of the experiment with 1024 executions using a quantum device simulator and the Rome quantum device. The Rome NISQ device has an output very close to the expected result. The circuit used to obtain this result is described in Fig.~\ref{fig:circ_exp}, where $c$ is a classical register. We remove the last CNOT of the controlled operation because the qubit 2 will be discarded. The resulting circuit has 10 CNOT operators because a quantum swap was necessary to run this circuit in the real quantum device with a limited qubit connectivity. The circuit used in the quantum device is described in Fig.~\ref{fig:experimentnisq}.

\subsubsection{Circuit depth}

The main difference between the divide-and-conquer state preparation and previous approaches is an exchange between circuit depth by circuit width. Table~\ref{tab:comp} presents the depth of the circuits generated using the proposed strategy, implementation of a version of~\cite{shende2006synthesis} available at~\cite{gadi_aleksandrowicz_2019_2562111} and a non optimized version of the algorithm described in~\cite{mottonen2005transformation}. The proposed strategy and~\cite{mottonen2005transformation} implementation are publicly available. The implementation of the proposed method shows its theoretical asymptotic time advantage to load a vector when the dimension is larger than 32. The proposed method has two main disadvantages: the linear number of qubits in relation to the logarithmic number in other methods, and the information entangled in the ancillary qubits.

\begin{table}[ht]
\centering
\begin{tabular}{|c|c|c|c|c|c|c|}\hline
n	&  dc depth	&  dc width  & 	 \cite{shende2006synthesis} depth 	&  \cite{shende2006synthesis} width & \cite{mottonen2005transformation} depth & \cite{mottonen2005transformation} width\\ \hline
4 &	 12 & 	 4 &	 3 &	 3 & 5 & 3\\
8 	& 31 &	 8  &	 17 &	 4 & 53 & 4\\
16  & 58 &	 16 &	 47 &	 5 & 277 & 5\\
32 & 93 &	 32 &	 105 &	 6 & 1237 & 6\\
64 & 136 &	 64 &	 239 &	 7 & 5205 & 7\\
128 &	 187 &	 128 &	 493 &	 8 & 21333 & 8\\
256 &	 246 &	 256 &	 982 &	 9 & & 9\\
512 &	 313 &	 512 &	 2025 &	 10 & & 10\\
1024 &	 388 &	 1024 &	 4009 &	 11 & & 11\\ \hline
\end{tabular}
\caption{A comparison between Refs. \cite{mottonen2005transformation}, \cite{shende2006synthesis} and divide-and-conquer strategy to load a $n$-dimensional real vector into a quantum computer.}
\label{tab:comp}
\end{table}

The higher depth of circuits using the divide-and-conquer strategy with small vectors occurs because of the use of three-qubits gates to combine the vectors. In other works, it is only necessary to use $O(n)$ qubits to load a $2^n$-dimensional vector while requiring sequential applications of $O(2^n)$ $n$-controlled gates. To improve the performance of the divide-and-conquer loading strategy and to reduce the number of qubits one can combine algorithm~\cite{shende2006synthesis} with the divide-and-conquer strategy. Instead of divide the vector in parts with size 2, we can divide the vector in parts with size $k$ (equal to a power of 2), load the normalized $k$-dimensional vectors using a sequential algorithm and combine the small vectors with the divide-and-conquer approach.

\subsection{Example Applications}

\subsubsection{Hierarchical Quantum Classifier}
\label{sec:hqc}
This section compares the divide-and-conquer algorithm with two other approaches in which input data encoding in a quantum state can be achieved to initialize a quantum circuit, namely qubit encoding and amplitude encoding. In the former, data is encoded in the amplitudes of individual qubits in a fully separable state, performed using single-qubit rotations \cite{grant_hierarchical_2018}. In the later, data is encoded in the amplitudes of an entangled state \cite{shende2006synthesis,mottonen_transformation_2005}, 
similarly to the divide-and-conquer. We use the accuracy of a quantum variational classifier as a metric to evaluate the state preparations. The divide-and-conquer algorithm is expected to produce results similar to the amplitude encoding. The results of the classifier using qubit encoding are also presented for completeness, albeit our main objective is to compare the divide-and-conquer and amplitude encoding schemes.

The classifier is based on a tree-like circuit known as tree tensor network (TTN) \cite{grant_hierarchical_2018}. This choice is based on the fact that tensor networks can represent both neural networks and quantum circuits, acting as a link between these fields \cite{cohen_convolutional_2016,levine_quantum_2019}. Initially, it applies a set of two-qubit unitaries to each pair of qubits from the initial state, discarding one output from each unitary, leaving half the number of qubits left for the next layer. The process is repeated until only one qubit remains. Multiple measurements are carried on this last qubit to approximate the expectation value.

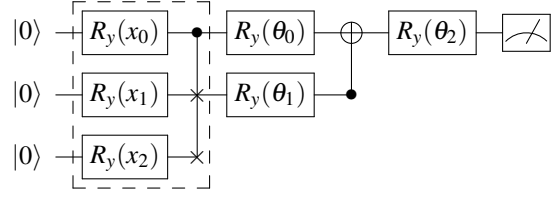
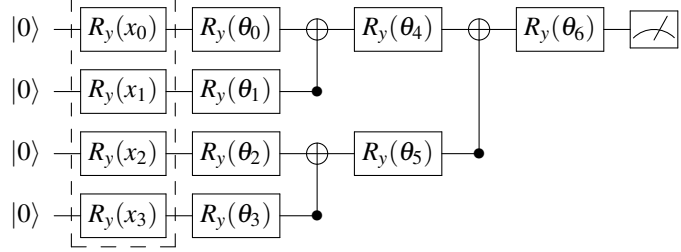
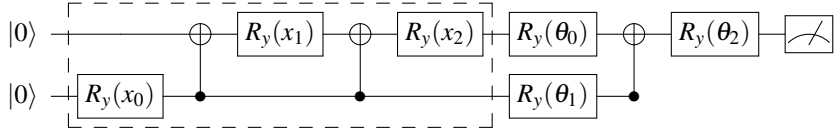
\begin{figure}[t]
\centering
    \begin{subfigure}[b]{\textwidth}
    $$
\Qcircuit @C=1em @R=.7em {
  \lstick{\ket{0}} & \gate{R_y(x_0)} & \ctrl{2} & \gate{R_y(\theta_0)} & \targ     & \gate{R_y(\theta_2)} & \meter \\
  \lstick{\ket{0}} & \gate{R_y(x_1)} & \qswap   & \gate{R_y(\theta_1)} & \ctrl{-1} &                      &        \\
  \lstick{\ket{0}} & \gate{R_y(x_2)} & \qswap   &                      &           &                      &        \\
  \protect\gategroup{1}{2}{3}{3}{.7em}{--}
}
$$
\caption{TTN classifier using divide-and-conquer encoding. Gates in the highlighted area encode each element of a length 3 classical data vector in the amplitudes of the entangled state composed by the first two qubits.}\label{Fig:dc_hqc}
    \end{subfigure}
    \begin{subfigure}[b]{\textwidth}
    $$
    	\Qcircuit @C=1em @R=.7em {
    	    \lstick{\ket{0}} & \gate{R_y(x_0)} & \gate{R_y(\theta_0)} & \targ     & \gate{R_y(\theta_4)} & \targ & \gate{R_y(\theta_6)} & \meter \\
    	    \lstick{\ket{0}} & \gate{R_y(x_1)} & \gate{R_y(\theta_1)} & \ctrl{-1} &                      & & & \\
    	    \lstick{\ket{0}} & \gate{R_y(x_2)} & \gate{R_y(\theta_2)} & \targ     & \gate{R_y(\theta_5)} & \ctrl{-2} & & \\
    	    \lstick{\ket{0}} & \gate{R_y(x_3)} & \gate{R_y(\theta_3)} & \ctrl{-1} &                      & & & \\
    	    \protect\gategroup{1}{2}{4}{2}{.7em}{--}
    	}
    	$$
    	\caption{TTN classifier using qubit encoding. Rotations in the highlighted area encode each element of a length 4 classical data vector in the amplitudes of individual qubits.}\label{Fig:qubit_hqc}
    \end{subfigure}
    \begin{subfigure}[b]{\textwidth}
    $$
	\Qcircuit @C=1em @R=.7em {
	    \lstick{\ket{0}} & \qw             & \targ     & \gate{R_y(x_1)} & \targ     & \gate{R_y(x_2)} & \gate{R_y(\theta_0)} & \targ &  \gate{R_y(\theta_2)} & \meter \\
	    \lstick{\ket{0}} & \gate{R_y(x_0)} & \ctrl{-1} & \qw             & \ctrl{-1} & \qw             &\gate{R_y(\theta_1)} & \ctrl{-1} &  & \\
	    \protect\gategroup{1}{2}{2}{6}{.7em}{--}
	}
	$$
	\caption{TTN classifier using amplitude encoding. Gates in the highlighted area encode each element of a length 3 classical data vector in the amplitudes of circuit's entangled state.\label{Fig:amp_hqc}}
    \end{subfigure}
    \caption{TTN classifier with \ref{Fig:dc_hqc} divide-and conquer encoding, \ref{Fig:amp_hqc} amplitude encoding and \ref{Fig:qubit_hqc} qubit encoding.}
\end{figure}

Following Grant et al. \cite{grant_hierarchical_2018}, we built the circuits using single-qubit rotations around the $y$-axis of the Bloch sphere, denoted by $R_y(\theta)$, and CNOT gates, composing two-qubit unitary blocks $CNOT\cdot (R_y(\theta_0) \otimes R_y(\theta_1))$. The single-qubit rotation angle $\theta$ is subject to training by some optimization procedure. Examples of the resulting circuits are represented in Figs. \ref{Fig:dc_hqc}, \ref{Fig:qubit_hqc}, and \ref{Fig:amp_hqc}.

We follow the general classical-quantum hybrid approach in which the optimization procedure is processed on a classical computer to determine a set of parameters, i.e. rotation angles for the $R_y(\theta)$ operation, for the parametrized quantum circuit. 
The quantum device prepares a quantum state as prescribed by the circuit pipeline and performs measurements. The measurement outcomes are processed by a classical device to generate a forecast, using it to update the model parameters via a learning algorithm. This whole process is repeated towards the goal. 

Four datasets were used in this work: Iris, Haberman's Survival, Banknote Authentication \cite{Dua:2019}, and Pima Indians Diabetes \cite{nr}. Three binary datasets were extracted from the original Iris dataset (paired combinations of the original three classes). Mean test accuracy and one standard deviation are computed on ten random initializations for each dataset and encoding. The simulation results are presented in Table \ref{Tab:accuracy}, where the test accuracy of the qubit and amplitude encodings are compared against the results obtained using the divide-and-conquer encoding.

\begin{table}[ht]
	\centering
	\begin{tabular}{|c|c|c|c|c|}
		\hline
		\multirow{2}{*}{Dataset} & \multirow{2}{*}{Classes} &\multicolumn{3}{c|}{Encoding} \\
		\cline{3-5}
		& & Qubit & Amplitude & Divide-and-conquer\\
		\hline
		\multirow{1}{*}{Haberman} & 0 or 1 & 60.33$\pm${2.02} & 59.02$\pm${0.00} & 59.02$\pm${0.00} \\
		
		\hline
		\multirow{1}{*}{Banknote} & 0 or 1 & 91.28$\pm${3.11} & 87.15$\pm${0.74} & 87.45$\pm${1.12}\\
		
		\hline
		\multirow{1}{*}{Pima}     & 0 or 1 & 77.19$\pm${2.08} & 70.78$\pm${1.88} & 71.11$\pm${1.79}\\
		
		\hline
		\multirow{3}{*}{Iris}     & 0 or 1 & 100$\pm${0.00}  & 100$\pm${0.00} & 100$\pm${0.00}\\ 
		
		\cline{2-5}
		& 0 or 2 & 100$\pm${0.00} & 100$\pm${0.00} & 100$\pm${0.00} \\
		
		\cline{2-5}
		& 1 or 2 & 98.50$\pm${2.42} & 93.00$\pm${2.58} & 93.00$\pm${2.58} \\
		
		\hline
		
	\end{tabular}
	\caption{Mean test accuracy and one standard deviation for TTN classifiers with ten different random parameter initializations. Three binary datasets were extracted from the original Iris dataset.}\label{Tab:accuracy}
\end{table}

The results show similar classification accuracy for all encodings, favoring qubit encoding due to the greater number of circuit parameters for the optimization. The main advantage of divide-and-conquer encoding over qubit encoding is the representation of encoded data in a quantum state of a reduced number of qubits, $\log_2 (N)$, compared to the initial state $N-1$. This also results in a lower depth classifier. Moreover, when the data is given by qubit encoding, TTN circuits can be evaluated efficiently using classical techniques~\cite{grant_hierarchical_2018}. This is not true when the input data is amplitude encoded. The advantage over amplitude encoding is a lower depth encoding circuit for $N\ge 64$ (Table \ref{tab:comp}).

To verify that the above comparison of the models is appropriate, a nonparametric statistical test was employed. We used the Wilcoxon paired signed-rank test \cite{Demsar:2006:SCC:1248547.1248548} with $\alpha=0.05$ to check whether there exist significant differences between the classification performances of compared encoders over the chosen datasets. As expected, we verified that amplitude encoding and divide-and-conquer encoding are statistically equivalent for all datasets.

\subsubsection{Swap Test}
Some metric between two data set encoded as $\sum_i x_i|i\rangle=\sum_i|\tilde{x}_i\rangle$ and $\sum_j y_j|j\rangle=\sum_j|\tilde{y}_j\rangle$ can be calculated with the state prepared by the divide-and-conquer state preparation and the swap test. The required state is
\begin{equation}
    |0\rangle\sum_{ij}|\tilde{x}_i\rangle|\tilde{y}_j\rangle|\psi_i\rangle|\phi_j\rangle,
\end{equation}
where $\sum_i|\tilde{x}_i\rangle|\psi_i\rangle$ and $\sum_j|\tilde{y}_j\rangle|\phi_j\rangle$ are prepared by the encoding scheme explained in Sec.~\ref{sec:orthonormalancillary} so as to make the ancillary states orthonormal.

After applying the swap test circuit to the above state, i.e. the Hadamard on the first (ancilla) qubit, the swap operation between the test register and the data register controlled by the ancilla qubit, and the Hamadard on the first qubit, one obtains
\begin{equation}
    \frac{1}{2}\left(|0\rangle\sum_{ij}\left(|\tilde{x}_i\rangle|\tilde{y}_j\rangle+|\tilde{y}_j\rangle|\tilde{x}_i\rangle\right)|\psi_i\rangle|\phi_j\rangle+
    |1\rangle\sum_{ij}\left(|\tilde{x}_i\rangle|\tilde{y}_j\rangle-|\tilde{y}_j\rangle|\tilde{x}_i\rangle\right)|\psi_i\rangle|\phi_j\rangle\right).
\end{equation}
Now, when the $\sigma_z$ measurement is performed on the ancilla qubit, the probability to measure $z=\pm 1$, i.e. $z=+1$ if the ancilla qubit is $|0\rangle$ and $z=-1$ if the ancilla qubit is $|1\rangle$, is
\begin{align}
   \Pr(z=\pm 1) &= \frac{1}{4}\sum_{ijkl}2\left(\langle \tilde{x}_k|\tilde{x}_i\rangle\langle \tilde{y}_l|\tilde{y}_j\rangle\pm\langle \tilde{y}_l|\tilde{x}_i\rangle\langle \tilde{x}_k|\tilde{y}_j\rangle\right)\langle\psi_k|\psi_i\rangle\langle\phi_l|\phi_j\rangle\nonumber \\
   & = \frac{1}{2}\left(\sum_{ij}\langle \tilde{x}_i|\tilde{x}_i\rangle\langle \tilde{y}_j|\tilde{y}_j\rangle\pm|\langle \tilde{y}_j|\tilde{x}_i\rangle|^2\right)\nonumber\\
   &=\frac{1\pm\sum_{ij}|\langle \tilde{y}_j|\tilde{x}_i\rangle|^2}{2}.
\end{align}
Therefore, measuring the expectation value of $\sigma_z$ on the ancilla qubit yields
\begin{equation}
\label{eq:swap_test_result}
    \sum_{ij}|\langle \tilde{y}_j|\tilde{x}_i\rangle|^2=\sum_{i} |x_iy_i|^2.
\end{equation}

Several measures in statistics can be derived from the above result. First, by setting $|x_i|^2$ to be the possible values of a discrete random variable $X:\Omega\rightarrow \mathbb{R}$ with the probability $\mathrm{Pr}(X=|x_i|^2) = |y_i|^2$, the above equation becomes an expectation value of the random variable $X$. The above equation can be also viewed as the second moment of a discrete random variable $X$, i.e. $E(X^2)$, with the probability $\mathrm{Pr}(X=x_i) = |y_i|^2$. This can be used to calculate the variance of $X$ given $E(X)^2$. Alternatively, the above equation can be viewed as $E(XY)$ of two uniformly-distributed discrete random variables $X$ and $Y$ that satisfy $\mathrm{Pr}(X=|x_i|^2)=\mathrm{Pr}(Y=|y_i|^2)=1/N$. This can be used with $E(X) = \sum_i^N |x_i|^2/N = E(Y) = \sum_i^N|y_i|^2/N=1/N$ to calculate the covariance, $E(XY)-E(X)E(Y)$.

The idea above can be extended for calculating the covariance of two discrete random variables $X$ and $Y$ with any known probability distribution. Let possible outcomes of $X$ and $Y$ be $(|x_0|^2,\ldots,|x_{N-1}|^2)$ and $(|y_0|^2,\ldots,|y_{N-1}|^2)$, respectively, and the probability distribution be $(p^{x}_0,\ldots,p^{x}_{N-1})$ and $(p^{y}_0,\ldots,p^{y}_{N-1})$, respectively. Then the divide-and-conquer algorithm can be used to prepare a state
\begin{equation}
    |0\rangle\sum_{ijkl}|\tilde{p}^x_i\rangle|\tilde{x}_j\rangle|\tilde{p}^y_k\rangle|\tilde{y}_l\rangle |\psi_{ijkl}\rangle,
\end{equation}
where $|\tilde{p}^x_i\rangle = \sqrt{p^x_i}|i\rangle$, $|\tilde{p}^y_k\rangle = \sqrt{p^y_k}|k\rangle$, $|\tilde{x}_j\rangle = x_j|j\rangle$, $|\tilde{y}_l\rangle = y_l|l\rangle$, and $|\psi_{ijkl}\rangle$ is the orthonormal ancillary state as before. Now, the swap test circuit is applied with a small modification such that $3n$ controlled-swap gates are applied to transform $|\tilde{p}^x_i\rangle|\tilde{x}_j\rangle|\tilde{p}^y_k\rangle|\tilde{y}_l\rangle$ to $|\tilde{x}_j\rangle|\tilde{p}^y_k\rangle|\tilde{y}_l\rangle|\tilde{p}^x_i\rangle$ when the ancilla qubit for the swap test is $|1\rangle$. Measuring the expectation value of the $\sigma_z$ observable on the ancilla qubit yields
\begin{equation}
    \sum_{ijkl}\langle \tilde{p}^x_i|\tilde{x}_j \rangle \langle \tilde{x}_j|\tilde{p}^y_k \rangle \langle \tilde{p}^y_k|\tilde{y}_l \rangle \langle \tilde{y}_l|\tilde{p}^x_i \rangle=\sum_{i} p^x_ip^y_i|x_i|^2|y_i|^2=E(XY).
\end{equation}
$E(X)$ and $E(Y)$ can be calculated from the swap test algorithm presented in the beginning of this section, which provided Eq.~\eqref{eq:swap_test_result}, by choosing the input vectors appropriately.

The total time complexity for the aforementioned quantum algorithms is still $O_q(\log_2^2(N))$, since the swap test only requires additional $O(\log_2(N))$ controlled-swap gates. The quantum speedup can be manifested when constructing a covariance matrix for two multivariate random variables $\mathbf{X}$ and $\mathbf{Y}$, each containing $m$ discrete random variables of size $N$. Since there are $m^2$ entries in the matrix, the classical time cost is $O_c(Nm^2)$, while the quantum approach requires $O_c(N) + O_q(\log_2^2(N) m^2)$.

\section{Discussion}
\label{sec:conclusion}

One of the major open problems for practical applications of quantum computing is to develop an efficient means to encode classical data in a quantum state~\cite{Aaronson_nature2020}. Most quantum algorithms do not present advantages in loading data~\cite{biamonte2017quantum}. The method proposed in this work fills this gap by proposing a new quantum state preparation paradigm, which can complement or enhance the known methods, such as qubit encoding and amplitude encoding. Our approach was based on the M{\"o}tt{\"o}nen et al. algorithm~\cite{mottonen2005transformation} and a divide-and-conquer approach using controlled swap gates and ancilla qubits. With this modification, we obtain an exponential quantum speedup in time to load a $N$-dimensional real vector in the amplitude of a quantum state with a quantum circuit of depth $O(\log_2^2(N))$ and space $O(N)$. The exponential speedup to load data in quantum devices has a potential impact on speeding up the solution of problems in quantum machine learning and other quantum algorithms that need to load data from classical devices. 

The speedup is achieved at the cost of using ancilla qubits that are entangled to the data register qubits. However, we showed that some interesting problems such as quantum supervised machine learning and statistical analysis can be performed with the input quantum state given by our method. The tradeoff between time and space complexities that our method provides is favorable when increasing the circuit width is easier than increasing the circuit depth, which is a likely scenario to occur during the development of near-term quantum devices.

We demonstrated the proof-of-principle using the IBM quantum cloud platform to verify the validity and the feasibility of our method. Furthermore, the numerical experiments showed that the new encoding method offers advantages, reducing complexity and computational resources when applied in conjunction with existing algorithms. Our perspective is that these advantages will extend to other cases.

This work leaves some open questions. What are other problems that can be solved with a divide-and-conquer quantum strategy? What are the implications to efficiently load a quantum vector with entangled information in the ancillary qubits for machine learning? And how to combine sequential with parallel strategies to create a robust algorithm with respect to input size? Also, finding an efficient means to uncompute the ancillary information remains as an interesting future work that will broaden the applicability of our method.

\section{Methods}

We performed the proof of concept experiment with a publicly available IBM quantum device consisting of five superconducting qubits, named as ibmq\_rome. The quantum circuit used in this experiment is depicted in Fig.~\ref{fig:circ_exp}. The circuit in Fig.~\ref{fig:circ_exp} is compiled to the physical qubit layout of ibmq\_rome and the resulting circuit is depicted in Fig.~\ref{fig:experimentnisq} that is executed 1024 times to obtain the data used to generate Fig.~\ref{fignisq}. We used the quantum information science kit (qiskit). Python implementation of \genangles and Algorithm~\ref{alg:dcload} are used to generate the quantum circuit in Figures \ref{fig:dccircuit} and \ref{fig:circ_exp}.

\begin{figure}[ht]
	$$
	\Qcircuit @C=0.37em @R=.4em {
	    \lstick{\ket{0}} & \gate{\underset{(0.927, 0, 0)}{U_3}} & \qw & \qw & \qw & \qw & \ctrl{1} & \qw& \qw & \qw & \ctrl{1} & \qw & \qw & \qw & \ctrl{1} &  \gate{\underset{(0.785)}{U_1}} & \ctrl{1}& \qw & \meter & \cw & \cw & \rstick{c[1]} \\
	    \lstick{\ket{0}} & \gate{\underset{(1.047, 0, 0)}{U_3}} & \ctrl{1}  & \gate{\underset{(0, 3.142)}{U_2}} & \targ     & \gate{\underset{(-0.785)}{U_1}} & \targ & \gate{\underset{(0.785)}{U_1}} & \targ & \gate{\underset{(-0.785)}{U_1}} & \targ & \ctrl{1} & \targ & \ctrl{1} & \targ &  \gate{\underset{(-0.785)}{U_1}} & \targ  \\
	    \lstick{\ket{0}} & \gate{\underset{(0, 2.220)}{U_2}} & \targ & \qw & \ctrl{-1} & \qw & \qw & \qw & \ctrl{-1} & \gate{\underset{(0.785)}{U_1}} & \qw & \targ & \ctrl{-1} & \targ  & \gate{\underset{(0, 3.927)}{U_2}} & \qw & \qw & \meter & \cw &\cw & \rstick{c[0]}
	}
	$$
	\caption{The transpiled circuit of the divide-and-conquer state preparation circuit in Fig.~\ref{fig:circ_exp} in accordance with the physical qubit layout of the ibmq\_rome device. The gates $U_1$, $U_2$, and $U_3$ are physical single-qubit gates of IBM Quantum Experience that take in one, two, and three parameters, respectively. The measurement outcomes are stored in classical registers denoted by $c[0]$ and $c[1]$.}
	\label{fig:experimentnisq}
\end{figure}

The depth of the quantum circuits for state preparations described in Table 1 is obtained using a python implementation of Algorithm~\ref{alg:dcload},
the qiskit implementation of \cite{shende2006synthesis} and a non-optimized version of the algorithm \cite{mottonen2005transformation} available at the GitHub repository. For each input size we generated a random vector used for all methods. In these first two set of experiments we used qiskit version 0.14.1 and python version 3.7.7.

In Section \ref{sec:hqc}, simulations of the hybrid classification algorithms were performed using Xanadu's Pennylane \cite{bergholm_pennylane_2020} default qubit plugin state simulator.
We used $2/10$ of the datasets as a test set, $2/10$ as a validation set, and the remaining as a training set. As preparation for qubit encoding, each data vector element of all datasets was re-scaled within the range of $[0, \pi]$. Also, for amplitude encoding and divide-and-conquer encoding, the data vectors were normalized. Our simulation employs the Adaptative Moment Estimation (Adam) for the optimization process \cite{kingma_adam:_2017} with a learning rate of $0.1$ and a batch size of $1/10$ of the training set size. Training stops when validation set accuracy does not increase for 30 consecutive tests or 200 iterations is reached.

\section*{Data availability}
The site \url{https://www.cin.ufpe.br/~ajsilva/dcsp/} contains all the data and software generated during the current study.

\bibliography{biblio}

\section*{Acknowledgements}

This research is supported by CNPq (Grant No. 308730/2018-6), CAPES – Finance Code 001 and FACEPE (Grant No. IBPG 0834-1.03/19), the National Research Foundation of Korea (Grant No. 2019R1I1A1A01050161 and 2018K1A3A1A09078001), the Ministry of Science and ICT, Korea, under an ITRC Program, IITP-2019-2018-0-01402, and the South African Research Chair Initiative of the Department of Science and Innovation and National Research Foundation (UID: 64812).
We acknowledge use of IBM Q for this work. The views expressed are those of the
authors and do not reflect the official policy or position of IBM or the IBM Q team.

\section*{Author Contributions}
AJS devised the divide-and-conquer state preparation strategy, performed the proof-of-concept experiments and wrote a first version of the manuscript. DKP suggested the way to make ancillary states orthogonal, and conceived the combination of swap test with the divide-and-conquer state preparation to calculate metrics between two datasets. IFA performed the experiments  with the variational quantum circuits and extended the algorithm to complex input vectors. All authors  reviewed and discussed the analyses and results, and contributed towards
writing the manuscript.

\section*{Competing interests}
The authors declare no competing interests.

\section*{Additional Information}
\textbf{Correspondence} and requests for materials should be addressed to F.P.

\end{document}